\documentstyle[11pt,fleqn]{article}
\def\ref{par\noindent\hangindent=6mm\hangafter=1}
\begin{document}
\vbox{
\rightline{Los Alamos electronic archives: gr-qc/9703051r}
}
\baselineskip 8mm
%
\begin{center}
{\bf Fisher's arrow of `time' in cosmological coherent phase space}

\bigskip

B. Roy Frieden$^{\dagger}$\footnote{E-mail: friedenr@super.Arizona.edu}
and H.C. Rosu$^{\ddagger}$\footnote{E-mail: rosu@ifug3.ugto.mx}

$^{\dagger}$
{\it Optical Sciences Center, University of Arizona, Tucson, Arizona 85721,
USA}

$^{\ddagger}$
{\it Instituto de F\'{\i}sica de la Universidad de Guanajuato, Apdo Postal
E-143, Le\'on, Gto, M\'exico}

\end{center}


\begin{abstract}

Fisher's arrow of `time' in a cosmological phase space
defined as in quantum
optics (i.e., whose points are coherent states) is
introduced as follows.
Assuming that the phase space evolution of the universe
starts from an initial squeezed cosmological state towards a final thermal
one, a Fokker-Planck equation for the time-dependent, cosmological
Q phase space probability distribution can be written down.
Next, using some recent results
in the literature, we derive an information arrow of time
for the Fisher phase space cosmological entropy based on the Q function.
We also mention the application of Fisher's arrow of time to
stochastic inflation models.

\end{abstract}
\bigskip
{\it PACS} number(s):  98.80.Hw 

\vskip 0.01cm
\begin{center}{\scriptsize Honorable Mention at Gravity Research Foundation
Essay Contest for 1997} 
\end{center}
\begin{center} {\scriptsize revised December 1997,
Mod. Phys. Lett. A 13 (1998) 39-46}
\end{center}

\vskip 0.01cm


\newpage


There has long been much interest in understanding the origin of
time-asymmetry of the Universe, i.e., the occurrence of definite
arrows of time \cite{book}. In this paper we give
arguments for introducing a Fisher arrow of `time' (where `time' is an
intrinsic time which is an unknown function of the foliation time, see below)
based on (i) the existence of a Fokker-Planck equation
in the phase space of cosmological coherent states for the $Q$ representation
of the density operator and
(ii) cosmological evolution starting from a squeezed state
to a final thermal one. Indeed,
quantum gravity and quantum cosmology (in the sense of quantum fluctuations
in the very early Universe) cannot escape quantum-optical phase space
approaches \cite{1},
which have proven so useful in both pure and applied physics ever since the
first phase space formulations of quantum mechanics have been given by Wigner
in 1932 and Husimi in 1940.
In the area of quantum optics there has been substantial progress precisely
as an outcome of the c-number phase space quasi-distribution functions
(representing a solution of the operator ordering problem for coherent states)
such as the $P$ function
introduced by Glauber and Sudarshan at the beginning of the sixties.
The seventies, and in
a much greater extent the eighties, brought
in the remarkable squeezed (non-classical) states \cite{sq},
of promising technological
potential in optical communication and high-precision interferometry. At the
same time, the squeezing formalism may be quite efficient in interpreting the
cosmological scales. This can be seen as follows. One way to produce
squeezed states is by parametric amplification due to the parametric
coupling with the pump field. On the other hand, in one of the first
contributions to quantum cosmology, Zel'dovich \cite{z1} pointed on the
analogy of cosmological particle production and the parametric
amplification of classical waves, and Grishchuk \cite{g1} actually suggested
the variable gravitational cosmic background acting similar to the pump
field in the laboratory squeezing. At present, the one-to-one correspondence
between the equations of the relic gravitons and those of quantum optics is
well known and it has been argued that some predictions of gravitation
theory can be tested in the laboratory by quantum optics experiments
\cite{ghb}. Although
Grishchuk and Sidorov \cite{gs} discussed plainly the cosmological
squeezing of the relic graviton spectrum, the squeezing approach
is not employed by many authors as yet \cite{lit}. This is partly due to the
opinion that the squeezing formalism is not promoting any new physics.
Despite this reticency, very recently,
Feinstein and P\'erez Sebasti\'an \cite{fps} went as far as describing the
birth of the universe out of ``nothing" in terms of quantum squeezing,
presenting the Casher-Englert tunneling entropy \cite{ce}
in the squeezing framework.

In the following, we shall pay particular attention to the $Q$ representation
of the cosmological density matrix, which is defined in terms of squeezed
coherent boson
states $|\alpha \rangle$. These quantum states are thought of as a
classical-like representation of the quantum fluctuations in the early
Universe and having well-defined amplitude and momentum are the best
analogues of points in the phase space \cite{hu}, which is considered to be
a locally compact, Hausdorff topological space.
They fulfill the decomposition of unity
$\int d^2\alpha|\alpha\rangle\langle \alpha|=\pi$ and lead to
$Q(\alpha)\equiv \frac{1}{\pi}\langle \alpha |\hat{\rho}|\alpha \rangle$.
Since the $Q$ function is positive by definition, can always be thought of
as a probability distribution on phase space, not as the Wigner function or
the Glauber-Sudarshan $P$ function, which are only quasi-distributions obeying
pseudo-Fokker-Planck (FP) equations, i.e., FP equations with a non-positive
definite diffusion matrix \cite{R}.
Moreover, the  $Q$ function has been shown to be the probability distribution
for the statistics in a particular model of simultaneous measurements of
position and momentum \cite{m}. Although the arguments in the following are
general and independent of any particular cosmological scenario, we mention
that the dispersion of these Gaussian states can be written as in \cite{s}
$$
\sigma ^{-2}(t(\tau))=
\frac{p}{2}\left(\frac{{\rm cosh} r(t(\tau))+
e^{2i\varphi (t(\tau))}{\rm sinh} r(t(\tau))}
{{\rm cosh} r(t(\tau))-e^{2i\varphi (t(\tau))}{\rm sinh} r(t(\tau))}\right)~,
\eqno(1)
$$
where $p$ is a model dependent parameter,
$t(\tau)$ (henceforth also $t$)
is an intrinsic time expressed as a so far unknown function
of the conformal time  $\tau$, $r$ is the squeeze parameter, and
$\varphi$ is the squeeze phase. Furthermore, in the case of
relic gravitons (see below),
Grishchuk and Sidorov \cite{gs} used Eq.~(1) with $t(\tau)=\tau$ and
derived the formula $p=\sqrt{n^2-1}-a^2$, where $n=2\pi R/\lambda$
is an index labeling the graviton modes representing nontrivial solutions
of a Schr\"odinger-like equation, and $a$ is the dimensionless
scale factor connected with the physical one, $R$,
by $a=(3\pi/2)^{1/2}(R/l_{p})$, with $l_{p}$ the Planck length.
The essential point now is that
the time evolution of the $Q$ function is governed by a Fokker-Planck equation
that can be obtained from the well-known FP equation
for the Glauber-Sudarshan $P$ function. It reads \cite{kw}
$$
\partial _{t}Q=\Bigg[\frac{\gamma}{2}
(\frac{\partial}{\partial \alpha}\alpha+
\frac{\partial}{\partial \alpha ^{*}}\alpha ^{*})+D\frac{\partial ^2}
{\partial \alpha \partial \alpha ^{*}}\Bigg]Q \equiv L_{FP}Q~,
\eqno(2)
$$
where $\gamma$ is the drift coefficient and $D$ is the diffusion one, which
turns out to be $\gamma(\bar{n}+1)$, where $\bar{n}$ is the mean number of
quanta. As we said the intrinsic time $t$ is a model dependent function of
the conformal time.
Since the
FP kinetic evolution is crucial for the arguments in the following
we add from the book of Davies \cite{D} some basic mathematical
structure that is implied by such an assumption.
Essentially, the requirement that the initial value problem be
well-posed implies that the FP solutions (as those of any other kinetic
equation) should define a Markov semigroup, i.e., a continuous, positivity
and normalization preserving one-parameter semigroup of operators on the
space of complex-valued continuous functions in phase space. This is a
mathematical terminology stemming from the fact that such semigroups are
in one to one correspondence with (time-homogeneous) stochastically
continuous Markov processes described by a transition probability
distribution, which is a probability Borel measure on the phase space.
With some further mathematical conditions (like the so-called Lindenberg
condition
on the transition probability distribution) it can be shown that the
generator of the Markov semigroup has the FP form with the diffusion and
drift coefficients related to the first two moments of the transition
probability distribution of the corresponding Markov process.
When mathematicians are speaking about  `time-homogeneous' stochastic
processes, by time they understand
an intrinsic, physical time that may be for example a Bohmian time \cite{bd}.
However, one can employ the Bohmian time only for rough estimates of the
semi-classical tunneling
time through a classically forbidden region of the cosmological scale
factor as in fact some authors did since in the same tunneling process
the conversion of a parametric (foliation) time variable into an intrinsic
time takes place \cite{bt,gh}. Thus, it appears that in order
to get the function $t(\tau)$ one should perform a detailed analysis of the
cosmological semi-classical tunneling.
Assuming all this as a background for the FP equation (2), it
can be applied to at least three cosmological scenarios as follows.

(i) {\it Gravitons}:
since the inflationary amplification
of the vacuum fluctuations leads to gravitons in a final ``squeezed vacuum"
quantum state, one can study the phase space evolution of the squeezed
vacuum towards the thermal equilibrium.

(ii) {\it Tunneling approaches}:
one can also consider the phase space evolution of the
squeezed state emerging after the tunneling out from ``nothing" \cite{fps}.

(iii) {\it Electromagnetic fluctuations}:
an entirely electromagnetic origin of the cosmic microwave anisotropy
recently detected by the Cosmic Background Explorer (COBE) \cite{cobe} has been
shown to be allowed by realistic stringy cosmologies \cite{ggv}. In such a
case, the cosmological coherent phase space is a more complicated form of the
quantum optical phase space.

Taking into account all these possibilities, we shall keep our discussion at
the general quantum optical level, since as for the case of relic gravitons,
the connection with various cosmological scenarios is merely a matter of
infering the function $t(\tau)$.
Assuming an initial squeezed state evolving to a thermal one,
a solution of Eq.~(2) has been obtained by Keitel and W\'odkiewicz in the form
\cite{kw} 
$$
Q_{s}
(\alpha, t)=\exp[a_{s}(t)|\alpha|^2+b_{s}(t)(\alpha ^2+\alpha ^{*2})
+c_{s}(t)\alpha +c_{s}^{*}(t)\alpha ^{*}+ N_{s}(t)]~,
\eqno(3)
$$
where the subscript $s$ stands for squeezing,
$n(t)=(\bar{n}+1)(1-e^{-\gamma t})$, while the rather cumbersome
expressions of the time-dependent functions $a_{s},b_{s},c_{s},N_{s}$
are given in \cite{kw} and will not be reproduced here.
An important property of
the FP evolution is to preserve the Gaussian character of the initial
state. This holds for any normalized Gaussian function of the type given in
Eq.~(3), with arbitrary $a,b,c,N$ not necessarily as given in \cite{kw},
fulfilling only the restrictions $a(t)<0$, $2|b(t)|<|a(t)|$ and
$N(t)$ determined from the normalization of $Q$.
It can be checked \cite{kw} that $Q_{s}(\alpha; \infty)\equiv Q_{thermal}=
\frac{1}{\pi(\bar{n}+1)}\exp[-\frac{|\alpha|^2}{(\bar{n}+1)}]$.

Let us pass now to the Fisher information defined as the volume integral
$I$ $\equiv \int
\frac{\nabla p\cdot \nabla p}{p}dV$ that can be introduced
for any system characterized by
a probability distribution $p$. It has been shown to provide
interesting insights in theoretical physics by one of the authors \cite{fr}.
This information quantity is the scalar trace of the
Fisher information matrix \cite{T}.
Arguments for a Fisher's information ``H-theorem" have been recently
advanced \cite{fs}, having been turned more sound by a direct proof due
to Plastino and Plastino \cite{pp}. We shall sketch in the following the
phase space analog of
Plastinos' proof for an arrow of time based on $I$. The adjoint operator
$L_{FP}^{+}$ is defined according to
$$
L_{FP}^{+}=
\Bigg[\left(\frac{\partial}{\partial \alpha}\alpha+
\frac{\partial}{\partial \alpha ^{*}}\alpha ^{*}\right)\frac{\gamma}{2}+
\frac{\partial ^2}{\partial \alpha \partial \alpha ^{*}}D\Bigg]~.
\eqno(4)
$$
Given two probability distributions $Q_{1}$ and $Q_{2}$, there exists the
relationship
$$
\int Q_1(L_{FP}Q_2)d^{2}\alpha=\int Q_2(L_{FP}^{+}Q_1)d^{2}\alpha ~.
\eqno(5)
$$
To any pair of solutions $Q_1$, $Q_2$, one associates the auxiliary quantity
$$
F=\int Q_1^2Q_2^{-1}d^2\alpha~,
\eqno(6)
$$
leading to the following relationship for its rate of change
$$
\frac{dF}{d\tau}=2\int q(L_{FP}Q_1)d^2 \alpha- \int Q_2(L_{FP}^{+}q^2)
d^2\alpha~,
\eqno(7)
$$
where $q=\frac{Q_1}{Q_2}$.
After some manipulations, one is led to the relationship
$$
\frac{dF}{dt}=-2\int Q_2D\frac{\partial q}{\partial \alpha}
\frac{\partial q}{\partial \alpha ^{*}}d^2 \alpha~.
\eqno(8)
$$
Since $Q_2$ and $D$ are positive quantities one gets
$$
\frac{dF}{dt}\leq 0~.
\eqno(9)
$$

Let us consider now a family of normalized probability distributions
$Q_{\theta}=Q(\alpha;t , \theta)$, depending upon a parameter
$\theta$, all of them being solutions of the FP equation (2).
Assuming $L_{FP}$ as a $\theta$-independent linear operator, we have
$$
\frac{\partial}{\partial t}
\left(\frac{\partial Q_{\theta}}{\partial \theta}
\right)=L_{FP}\left(\frac{\partial  Q_{\theta}}{\partial \theta}\right)~.
\eqno(10)
$$
In this way, $\frac{\partial Q_{\theta}}{\partial \theta}$ is itself a
solution of the FP equation (not necessarily a normalized,
positive-definite one). Using the definition of Fisher's information
$I_{\theta}$ as follows
$$
I_{\theta}=\int \frac{1}{Q_{\theta}} \left(\frac{\partial Q_{\theta}}
{\partial \theta}\right)^2 d^2 \alpha
\eqno(11)
$$
and substituting $Q_1$ by $\frac{\partial Q_{\theta}}{\partial \theta}$ and
$Q_2$ by $Q_{\theta}$ in equation (6), comparison
with equations (8) and (9) leads to
$$
\frac{dI_{\theta}}{dt}\leq 0~.
\eqno(12)
$$

This result may be called the `I-theorem', by analogy with Boltzmann
H-theorem. It states that the Fisher information of the universe
(actually of any physical system obeying a FP equation) can
only decrease (or remain constant) in physical time.

A simplified proof \cite{fr2} is based on the fact that Fisher $I$ is
actually the cross-entropy between $Q(\alpha)$ and $Q(\alpha +d\alpha)$.
It is known, see e.g., Risken's book \cite{R}, that the cross-entropy obeys
$\frac{dI}{dt}\leq 0$, if $Q$ obeys a true (i.e., the diffusion
coefficient nonnegative)  FP equation.
This is independent of the parameter $\theta$ of the other proof.

The aforementioned applications of Fisher's arrow of `time' to various
cosmological scenarios are not
the only ones. In an even more direct way, `I-theorem'-like statements can
be easily introduced in the stochastic approach to models of eternal
inflation, as discussed for example by Winitzki and Vilenkin \cite{stoch}.
These models are based on FP equations for an inflaton
probability distribution $P(\phi, t)d\phi$ which is interpreted as the (either
comoving or physical) volume of
regions with a particular value of the inflaton scalar field in the
infinitesimal interval $d\phi$. Winitzki and Vilenkin employed a time variable
$t$, which is related to the proper time $\tau$ by
$$
dt=[H(\phi)]^{1-\alpha}d\tau~,
\eqno(13)
$$
where $\alpha$ is called the physical dimension of $t$, taking the value
$\alpha =1$ for the proper time parametrization, and $\alpha =0$ for the
scale time; $H(\phi)$ is the Hubble inflaton constant, proportional to the
square root of the inflaton potential.

On the other hand, when introducing Fisher's
arrow of `time' in Wheeler-DeWitt quantum cosmologies one is faced with the
ambiguity of defining a time variable, or in other words,
with the famous problem of time \cite{bp} (not to
mention the related issue of the emergence of time), which
is a truly fundamental
problem not only of cosmology but of physics in general.
So the question is: Does an arrow of time make any sense for the common
quantum
cosmological framework based on the Wheeler-DeWitt equation, which being
a stationary, zero energy differential equation appears to have no obvious
time dependence and therefore forcing one to invent clock models.
The point is that introducing an arrow of `time' means merely introducing
irreversibility in quantum cosmology. It has been suggested to us by the
Referee that irreversibility in a rigorous sense is given by an
underlying semigroup structure of the Lax-Phillips scattering type
\cite{LP}.
This proposal is reasonable and can be matched to the Wheeler-DeWitt
equation only if one thinks of the `wavefunction of the universe' solution
as similar to a zero-energy resonance (also known as a half-bound state,
and possible only for zero angular momentum) in ordinary quantum
scattering. The zero-energy resonance wavefunction is zero at
the arbitrary chosen cosmological origin, is finite at infinity, and is
not normalizable. Therefore, scattering concepts can be applied to it
\cite{ros}.

Coming back to the `I-theorem', another interesting byproduct is provided
by the
Cram\'er-Rao inequality for the mean square error $e^2$ of an unbiased
estimator
$e^2\geq I^{-1}$, indicating that even in the presence of efficient estimation
(i.e., Cram\'er-Rao equality)
{\em the quality of estimates must decrease with the time parameter}.
This seems to be a
reasonable alternative way of stating the second law of thermodynamics in
this context \cite{rfp}. This conclusion refers to processes for which the
necessary ingredients to the FP equation are fulfilled \cite{R},
i.e, they should be of zero
memory and obeying time reversal (only those within classical mechanics or
quantum mechanics without spin).

 After the completion of this work, we learned that
Rothman and Anninos \cite{ra} discussed
a phase space approach (volumes below the Hamiltonians of the systems they
study) to the ``gravitational arrow of time" (which by definition is
pointing in the direction of increased inhomogeneity).

{\bf Acknowledgment}

We thank the Referee for suggestions, especially the Lax-Phillips argument.
H.C. Rosu acknowledges partial support from CONACyT through the Project
4868-E9406.


\end{document}